# Output Privacy Protection With Pattern-Based Heuristic Algorithm

P. Cynthia Selvi[1] and A.R.Mohammed Shanavas[2]

[1]Associate Professor, Dept. of Computer Science, KNGA College(W), Thanjavur 613007 /Affiliated to Bharathidasan University, Tiruchirapalli, TamilNadu, India.
[2]Associate Professor, Dept. of Comuter Science, Jamal Mohamed College, Tiruchirapalli 620 020/ Affiliated to Bharathidasan University, Tiruchirapalli, TamilNadu, India

## ABSTRACT

*Privacy Preserving Data Mining(PPDM) is an ongoing research area aimed at bridging the gap between the collaborative data mining and data confidentiality There are many different approaches which have been adopted for PPDM, of them the rule hiding approach is used in this article. This approach ensures output privacy that prevent the mined patterns(itemsets) from malicious inference problems. An efficient algorithm named as Pattern-based Maxcover Algorithm is proposed with experimental results. This algorithm minimizes the dissimilarity between the source and the released database; Moreover the patterns protected cannot be retrieved from the released database by an adversary or counterpart even with an arbitrarily low support threshold.*

## 1. INTRODUCTION

Modern organizations manage a large amount of data, which can be mined to generate valuable knowledge, using several available data mining techniques. While data mining is useful within an organization, it can yield further benefits when it handles the combined data of multiple organizations. Many potential privacy problems are created by sharing of data.

The research on PPDM, are motivated by many of the applications in previous publications. In [1] the authors discussed the confidentiality of sensitive information in various dimensions and they emphasized that data mining technology provides a whole new way of exploring large datasets; but they may create problems in the wrong hands. How various data mining techniques can be used to provide business competitors an advantage in a business settings have been stated by the authors.

In this work, a pattern-based sanitization algorithm is suggested which is a heuristic technique to hide only selected values rather than all available values in large dataset that minimize the utility loss. The task of transforming the source dataset into a new dataset that hides some sensitive patterns or rules is called the sanitization process[2]. Sanitization procedures act on the data to remove or hide, or even alter sensitive information from the data. This study aims at reducing the accidental hiding of non-sensitive patterns and the difference between the source and the new sanitized dataset.

## 2. RELATED WORK

Rule hiding research in PPDM focuses on the process of modifying the original dataset in such a way that certain frequent itemsets or sensitive association rules are transformed without seriously affecting the infrequent itemsets and non-sensitive rules. Basically, association rules provide a simple and useful form of patterns for data mining; once the frequent patterns(itemsets) are known, association rules can be very easily formed. A wide ranging survey on association rule hiding in [3] divides association rule hiding methods as exact approaches, border based and heuristic.

Exact approaches [4, 5] give optimal solution and have no side effect but have much computational cost. Border based approaches uses border theory [6, 7, 8, 9] in which the elements on the border are boundary to the infrequent itemsets. During hiding process, sensitive frequent itemsets are focused on preserving the quality of the border. Heuristic approaches uses heuristics for modifications in the dataset, which are efficient, scalable and fast. These are based on decreasing support and confidence. There are two types of techniques called distortion and blocking. Distortion techniques select transactions which hold sensitive itemsets and then delete the selected items from transactions to modify the dataset. This approach is based on support reduction; a framework for frequent itemset hiding is proposed in [10]. In Blocking items are replaced with unknown values instead of deleting the items in order to modify dataset. In [11] blocking is used and the idea behind this approach is to replace false values but this approach may have bad consequences.

Many association rule hiding algorithms are Apriori[12]-based that needs multiple scans of database or FP tree [13], which is unrealistic to construct main-memory based FP tree when the database is large. But in this work Matrix-Apriori[14] is used for finding frequent itemsets which is to be done prior to data sanitization; it works without candidate generation and uses a simpler data structure and shown to be faster. In [15], Yildz propose an algorithm which makes use of Matrix-Apriori approach but his approach deals only with mutually exclusive patterns.

## 3. PPDM TERMINOLOGY

### 3.1. Preliminaries

*Transactional Database: A* transactional database consists of a file where each record represents a transaction that typically includes a unique identity number (*trans_id*) and a list of items that make up the transaction.

*Association Rule :* It is an expression of the form $X \Longrightarrow Y$ , where X and Y contain one or more patterns(categorical values) without common elements.

*Frequent Pattern:* A pattern(itemset) that forms an association rule is said to be frequent if it satisfies a prespecified minimum support threshold.

*Privacy Preservation:* The model of privacy preservation is based on the removal of selective items in the restrictive patterns that are usually decided by the user, who may either be the owner or the contributor of the data.

### 3.2 Definitions

*Source Database:* A source database contains a set of all transactions, T in which each transaction containing itemset $X \in D$. In addition, each k-itemset $X \subseteq I$ has an associated set of transactions $T \subseteq D$, where $X \subseteq t$ and $t \in T$.

*Restrictive Patterns:* Pattern to be hidden from the transactional source database according to some privacy policies.

*Sensitive Transactions:* A transaction is said to be sensitive, if it contain atleast one restrictive pattern.

*Transaction Size:* The number of items that forms a transaction is the size of the transaction.

*Transaction Degree:* The degree of a sensitive transaction is defined as the number of restrictive patterns which it contains.

*Cover:* The *Cover*[16, 17, 18] of an item $A_k$ can be defined as,

$C_{Ak} = \{ \ rp_i \ / \ A_k \in rp_i \subset R_P, \ 1 \leq i \leq |R_P| \}$

*i.e.,* set of all restrictive patterns which contain $A_k$. The item that is included in a maximum number of $rp_i$'s is the one with *maximal cover or maxCover;*

*i.e., maxCover = max( |C$_{A1}$|, |C$_{A2}$| , ... |C$_{An}$| )* such that $A_k \in rp_i \subset R_P$.

# 4. SANITIZATION ALGORITHM

Given the source *dataset(D),* and the *restrictive patterns(R$_P$)*, the sanitization process tries to protect $R_P$ by decreasing the support values of restrictive patterns by removing items from sensitive transactions against the mining techniques used to disclose them. This process mainly includes four subtasks:

1. identify the set of sensitive transactions for each restrictive pattern;
2. select the (partial) sensitive transactions to sanitize;
3. identify the *victim item* to be removed;
4. rewrite the modified database after removing the *victim items*.

Basically, all sanitizing algorithms differs only in subtasks 2 & 3. In this section, the proposed sanitizing algorithm and the heuristics to sanitize a source database are introduced.

***Heuristic-1:*** To solve subtask-2, $S_T$ is sorted in decreasing order of *(deg + size)* which enables multiple patterns to be sanitized in a single iteration. The number of sensitive transactions to be sanitized for each restrictive pattern $rp_i \in R_P$ is controlled by the *victim transactions* (ie the transactions which do not decrease the support count of $rp_i$ in successive iterations).

***Heuristic-2:*** To solve subtask-3 the following heuristic is used:

Find T = $\bigcap_{i=1}^{|RP|} t\_list(rp_i)$. For every $t \in$ T, select the *victim item* $A_k$ with *maximal cover* such that $A_k \in rp_i \subset t$ and remove; In case of tie, choose one in *round robin*; continue selecting the victim items on this strategy for the left over transactions until the support count of all $rp_i$ becomes zero.

## 4.1. Proposed Algorithm

***Pattern-based MaxCover Algorithm(PMA):*** // based on Heuristics 1 & 2 //
Begin
Step 1 : a)  calculate *supCount(rp$_i$)* $\forall$ $rp_i \in R_P$ and sort in decreasing order ;
       b)  find *cover*  for every item $A_k \in R_P$ and sort in decreasing order ;
Step 2 : a)  extract *Sensitive Transactions-S$_T$* w.r.t. $R_P$ ;
       b)  find *deg(t), size(t)* $\forall$ $t \in$ $S_T$ ;
       c)  sort $t \in$ $S_T$ in decreasing order of *deg & size* ;
Step 3 :  filter  $\sim S_T \leftarrow$ D $-$ $S_T$ ;          // $\sim S_T$ - non sensitive transactions //

Step 4 : // Find $S_T$'-sanitized transactions //
*Module-i:*

    find T = $\bigcap_{i=1}^{|RP|} t_{list(rp_i)}$;

    for every $t \in T$ do

    {

        delete $A_k$ with *maxCover* such that $A_k \in rp_i \subset t$ (*round robin* in case of tie);

                                        // $A_k$ –*victimItem* //

        decrease *supCount* of all $rp_i$ which contain *victimItem* ;  // ie, $A_k \in rp_i \subset t$ //

        mark $t$ as *victimTransaction* w.r.t each $rp_i$ ;

    }

*Module-ii:*

    for every $rp_i \in R_P$ do

    {

        if (supCount < > 0)

        {

            for each $t \in$ *nonVictimTransactions* do

            {

                delete item $A_k$ with *maxCover* (*round robin* in case of tie);

                decrease *supCount* of every $rp_i$ ;  // ie, $A_k \in rp_i \subset t$ //

            }

        }

    }

Step 5 : //merge sanitized transactions with nonsensitive ones//

    D' $\leftarrow$ ~ $S_T \cup S_T$'

End.

## 4.2. Illustration

The algorithm given above is based on the Heuristics 1 and 2. In module-i, the sensitive transactions which are common for all restrictive patterns are identified[T = $\bigcap_{i=1}^{|RP|} t_{list(rp_i)}$], in order to speed up the sanitization process. For every transaction $t$ in T, delete $A_k$ with *maxCover*. Suppose more than one item exists, delete item in *round robin* fashion for successive transactions. Decrease the *supCount* of all *restrictive patterns($rp_i$)* which contain $A_k$ and repeat the process till the *supCount* of all restrictive patterns becomes zero. Thus an adversary could not trace the restrictive patterns with any arbitrary *minSup*.

This algorithm make use of some preprocessed lookup tables, which link the restrictive patterns and their items with the corresponding transactions that reduces the execution time. Also the victim transactions are identified with a lookahead procedure such that a transaction marked with a victim item for a particular restrictive pattern would not be considered again in the next iteration for any other item of the same pattern(which is of no use in decreasing the *supCount*).

***Example :*** The proposed Heuristics work is illustrated by the following example. Refer the *Source Dataset(D), Restrictive Patterns(Rp)* and *cover of restrictive items* given below.

Table.1 Source Database

| Tid | Itemset |
|-----|---------|
| t1 | a,b,c,d,e |
| t2 | a,c,d,f |
| t3 | c,e,f |
| t4 | c,b,e |
| t5 | a,b,c,d,f |

Table.2 Restrictive Patterns-R_P

| Rid | Pattern | SupCount |
|-----|---------|----------|
| r1 | a,c | 3 |
| r2 | c,d | 3 |
| r3 | d,f | 2 |

Table.3 Cover of Restrictive Items

| Ite | Pattern | Cover |
|-----|---------|-------|
| c | r1, r2 | 2 |
| d | r2, r3 | 2 |
| a | r1 | 1 |
| f | r3 | 1 |

Here, [*t2, t5*] are common for all $rp_i$'s. In both *t2* and *t5*, items *c* and *d* have *maxCover* (ie, 2). Hence, remove *c* and *d* by *round robin* fashion in *t2* and *t5* respectively. Removing *c* in *t2* decrease the *supCount* of *r1* and *r2*. Mark *t2* as *victim transaction* for *r1* and *r2*; because, further consideration of *t2* for *r1* and *r2* has no effect in decreasing their *supCount*. Similarly, removing *d* in *t5* decreases the *supCount* of *r2* and *r3*. Mark *t5* as *victim transaction* for *r2* and *r3*.

As this approach decrease the *supCount* of all $rp_i$ to 0, module-ii is included to consider the $rp_i$'s whose *supCount* is not equal to zero; for every transaction *t* which include $rp_i$, remove item $A_k$ with *maxCover*; If more than one item exists, choose one in *round robin* for successive transactions. Decrease the *supCount* of all *restrictive patterns( $rp_i$)* in *t* which contain $A_k$. The set of modified sensitive transactions are denoted as $S_T$'. Then the Sanitized database *D'* is formed by merging $\sim S_T$ and $S_T$'.

## 5. EXPERIMENTAL RESULTS

The experimental analysis was performed to evaluate the effectiveness and efficiency of the proposed algorithm; which were conducted on a system with AMD Turion-II N550 - Dual core processor - 2.6 GHz - 2GB RAM - 32 bit OS - windows 7 - Netbeans 6.9.1. - SQL 2005. Contiguous logical segments of real datasets *Retail(D1)* and *T10I4D100K(D2)* [19] were used whose characteristics are given in Table-4 & Table-5.

Table-4 : Characteristics of Source Dataset(D)

| Dataset Name | No. of Trans. | No. of Distinct Items | Min Len. | Max Len. | Min. no. of Sensitive Trans. | Max. no. of Sensitive Trans. | Dataset Size (MB) |
|---|---|---|---|---|---|---|---|
| Retail | 1K – 8K | 3176 – 8126 | 1 | 58 | 22 | 2706 | 90.2KB –729KB |
| T10I4D100K | 1K – 8K | 795 - 862 | 1 | 26 | 7 | 78 | 94.8KB –699 KB |

Table-5 : Characteristics of Restrictive Patterns(Rp)

| Dataset Name | No. of Restrictive Patterns | MinSup (%) | MaxSup (%) | MinConf (%) | Maxconf (%) | Min. Len. | Max.Len. |
|---|---|---|---|---|---|---|---|
| Retail | 5 – 25 | 0.8 | 32 | 16.8 | 91.7 | 2 | 5 |
| *T10I4D100K* | 5 – 25 | 0.6 | 5 | 32.5 | 85.7 | 2 | 6 |

Frequent patterns were obtained using Matrix Apriori approach. Once the Restrictive Patterns(Rp) are identified(based on some decision making policies) among the frequent patterns, following lookup tables were constructed:

*Lookup table-I :* A hash table that link all *restrictive items* to the associated transactions, which is used to
-   obtain the transactions associated with each *restrictive pattern($rp_i$)*, by finding intersection of corresponding row entries of the items in $rp_i$.

*Lookup table-II :* A hash table that link all *restrictive patterns* to the associated transactions, which is used to
-   obtain the *sensitive transactions*, by finding the union of all cell entries;
-   find the *supcount* of each $rp_i$ in Rp by counting the corresponding row entries.
-   find the *degree* of transactions by counting the corresponding column entries

*Lookup table-III :* A hash table that link each *restrictive item* to the associated *restrictive patterns* which is used to
-   obtain the *cover* of each *restrictive item* by counting the corresponding row entries.

All these preprocessing work requires time ranging between $1.75 - 9.02$(for $R_P$ with 5-25 in No. & transactions 1k-8k in No.) minutes and it reduces the execution time of the proposed sanitization algorithm. However, the entire sanitization work is to be done offline and hence it does not affect the process significantly. Moreover no exclusive search and sort procedures are used; instead simple queries are used to extract the required information from the hash tables constructed, which also speed up the entire process and the source database is not at all scanned again.

## 5.1. Performance Evaluation:

It refers to the task of executing correct program on various datasets and determining how much computing time and storage an algorithm require. The proposed algorithm is evaluated based on the performance metrics suggested in [10], which quantifies the difference between the source dataset and the released(sanitized) one.

The proposed algorithm was tested for the performance metrics on the basis of criteria stated below and the graphs for the observed readings are shown towards the end of this section.
Criteria-I : Number of *Restrictive Patterns* ranging between 5 and 25 (randomly chosen in 1K transactions) were selected and the results are shown in *Table-6.*

Criteria-II : Number of Transactions ranging between 2K and 8K (with 5 randomly chosen *Restrictive Patterns*) were used and the results are shown in *Table-7.*

## 5.2. Measures of Effectiveness

*Hiding Failure(HF):* This metric is measured in terms of percentage of restrictive patterns that are discovered from the released(sanitized) dataset(D') and it is given by,

$$HF = \frac{|RP(D')|}{|RP(D)|},$$

where $|RP(X)|$ denotes the number of restrictive patterns discovered from dataset X.

In this work, the proportion of restrictive patterns that may be discovered from the sanitized dataset(D') is *completely zero*. Moreover, the restrictive patterns cannot be discovered with any arbitrary low frequency threshold values.

*Misses Cost(MC):* As there are functional dependencies between restricted and non-restricted patterns, some rules would be removed accidentally, which may happen when some of the non-restrictive patterns lose support in the dataset during sanitization process. It is measured by,

$$MC = \frac{|\sim RP(D)| - |\sim RP(D')|}{|\sim RP(D)|},$$

where $\sim|RP(X)|$ denotes the number of non-restrictive patterns discovered from dataset X. The misses cost of this algorithm is observed to be 0% - 2.72%.

*Sanitization Rate(SR):* The ratio of removed items(*victim items*) to the total support count of restrictive patterns($rp_i$) in the source dataset D is termed as sanitization rate which is given by,

$$SR = \frac{|victim\ items|}{\sum_{i=1}^{k} supCount(rpi)}$$

As the number of items removed is minimal in this approach compared to the total number of items in the restricted patterns, the SR is observed to be nominal(40.75%-72.46%) and it is also interesting to note that SR linearly decreases when the number of patterns to be hidden are increased.

*Artifactual Pattern(AP):* AP occurs when some artificial patterns are generated in D' as an outcome of the sanitization process. It is given by,

$$AP = \frac{|P'| - |P \cap P'|}{|P'|},$$

where $|RP(X)|$ denotes the number of artificial patterns generated from dataset X. As the proposed algorithm hides restrictive patterns by removing items instead of swapping and replacement, from the source dataset(D), it does not generate any artifactual patterns.

## 5.3. Measures of Efficiency

*Dissimilarity(dif):* The dissimilarity between the original(D) and sanitized(D') databases is measured by comparing their contents instead of their sizes and it is calculated by,

$$dif(D, D') = \frac{1}{\sum_{i=1}^{n} fD(i)} x \sum_{i=1}^{n} [fD(i) - fD'(i)]$$

where $fx(i)$ represents the $i^{th}$ item in the dataset X. The proposed approach has very low percentage(0.22%-5.21%) of dissimilarity for all criteria;

*Execution Time:* The execution time and the scalability of the proposed algorithm is obtained by varying status of the restrictive patterns(various strategies), the number of patterns to be hidden and the size of the dataset. It is observed that the execution n time linearly increases for the increase of number of restrictive patterns and also the size of the dataset. This scalability is mainly due to the Lookup Tables used for indexing the sensitive transactions per restricted items and patterns. Moreover, there is no need to scan the source dataset again whenever we want to access a transaction during sanitization process; Thereby this algorithm outperforms the previous approaches, all of which require more than one scan.

Table.6. Results of Criteria-I

| | Dataset | No. of Restrictive Patterns | | | | |
|---|---|---|---|---|---|---|
| | | 5 | 10 | 15 | 20 | 25 |
| HF (%) | D1 | 0 | 0 | 0 | 0 | 0 |
| | D2 | 0 | 0 | 0 | 0 | 0 |
| MC (%) | D1 | 0 | 2.27 | 2.37 | 2.52 | 2.72 |
| | D2 | 0.18 | 0.37 | 0.31 | 0.80 | 1.14 |
| SR (%) | D1 | 72.46 | 71 | 71.18 | 71.25 | 69.44 |
| | D2 | 40.75 | 40.37 | 40.78 | 43.54 | 45.68 |
| Dif (%) | D1 | 2.35 | 3.92 | 4.39 | 4.81 | 5.21 |
| | D2 | 0.32 | 0.44 | 0.72 | 0.90 | 1.10 |
| Time(secs) | D1 | 5.79 | 5.57 | 5.86 | 7.46 | 8.25 |
| | D2 | 3.49 | 6.32 | 7.49 | 9.91 | 10.54 |

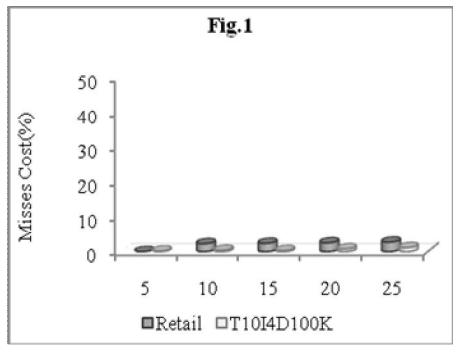
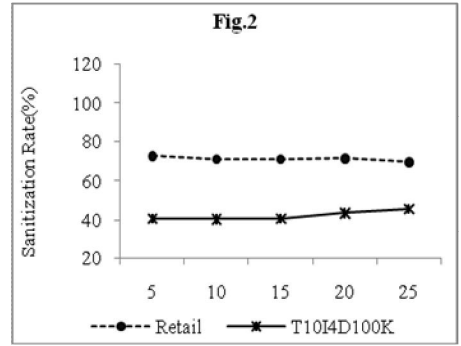

Fig.1&2.Measures of Effectiveness for varying number of patterns (criteria-I)

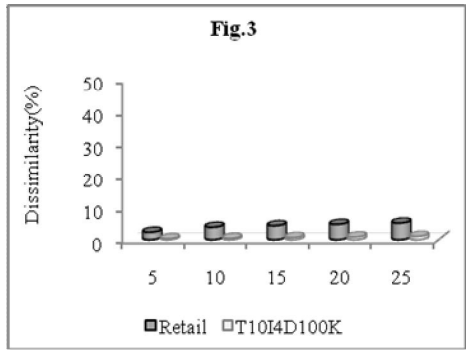
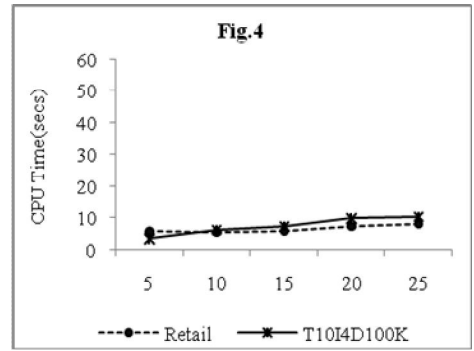

Fig.3&4.Measures of Efficiency for varying number of patterns (criteria-I)

Table.7. Results of Criteria-II

| | Dataset | No. of Transactions | | | |
|---|---|---|---|---|---|
| | | 2k | 4k | 6k | 8k |
| HF (%) | D1 | 0 | 0 | 0 | 0 |
| | D2 | 0 | 0 | 0 | 0 |
| MC (%) | D1 | 0 | 0 | 0 | 0 |
| | D2 | 0.94 | 0.1 | 0.09 | .14 |
| SR (%) | D1 | 69.18 | 69.29 | 70.79 | 68.33 |
| | D2 | 62 | 63.19 | 63.23 | 64.16 |
| Dif (%) | D1 | 0.56 | 0.43 | 0.4 | 0.35 |
| | D2 | 0.31 | 0.25 | 0.3 | 0.22 |
| Time (secs) | D1 | 3.68 | 4.71 | 6.93 | 5.18 |
| | D2 | 3.41 | 4.49 | 5.51 | 5.35 |

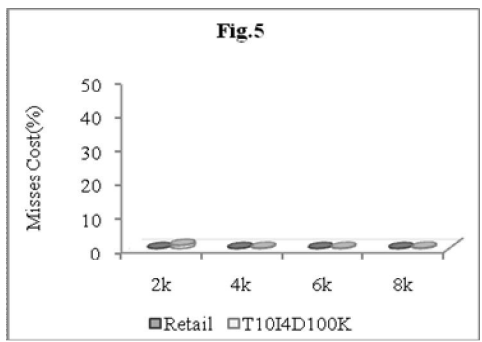
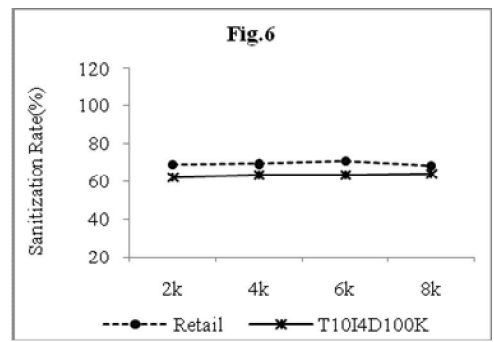

Fig.5&6.Measures of Effectiveness for varying number of transactions (criteria-II)

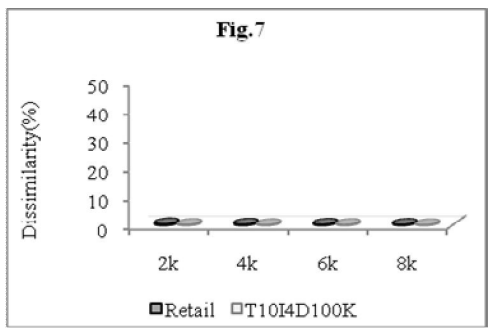
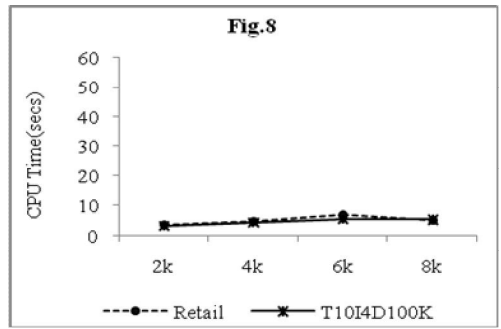

Fig.7&8.Measures of Efficiency for varying number of transactions (criteria-II)

## 5.4. Scalability :

To effectively hide information from a huge dataset, the sanitization algorithms must be efficient and scalable. That means the running time of a sanitizing algorithm must be predictable and acceptable in large datasets. The efficiency and scalability of this *PMA* approach is proved below:

*Theorem : The running time of the Pattern-based Maxcover Algorithm is $O(n \times N)$ in the worst case, where n is the number of restrictive patterns($rp_i$) and N is the number of transactions in the source dataset.*

*Proof :* Let *X* be a transactional dataset, *N* the number of transactions in X, *n* the number of restrictive patterns($rp_i$), in *X*.

In Step-1a, the *supCount* of each $rp_i$ is obtained from the LookUp Table-I & II ( hash tables that links each *restrictive item* and *restrictive patterns* with its associated transactions) and sorted in decreasing order using a query statement; In Step-1b, the *cover* of every restrictive item is found using LookUp Table-III (a hash table that links each *restricted item* with its associated $rp_i$ ) and sorted in decreasing order using a query statement; Similarly the statements in Step-2 and Step-3 are performed using LookUp Table-II, using queries. As all these steps involve straightforward computations and they take *O(1)* meaning that the computing time is a constant [20].

In Step-4, module-i requires *O(N)* in the worst case; and module-ii requires *O(n × N)*. All statements in module-i & ii are straightforward computations that are facilitated by simple query statements. Even the round robin selection of *victim items* are enabled by array indices. When there is a situation that both modules are to be performed (because in *PMA* approach the *supcount* of each $rp_i$ is reduced to zero), the execution time complexity is *O(N + n × N)*. Considering that $N << n \times N$, the running time of the main module is *O(n × N)*.

Again, Step-5 is a straightforward computation that merges the non-sensitive($\sim S_T$) and sanitized($S_T$) transactions using query and it requires *O(1)*.

Hence, the running time of the proposed *PMA* approach in the worst case is *O(n × N)* which is *linear* and better than $O(n^2)$, $O(n^3)$, $O(2^n)$, $O(n \log N)$. This linearity in execution time is also proved in the experimental results.

## 6. Conclusion

In order to facilitate data sharing with privacy protection in the cooperative business scenario, the algorithm *PMA* is proposed; This approach decrease the support count of maximum number of sensitive patterns, with possibly minimum number of removal of items. This strategy reduce the impact on the source dataset for privacy preserving in mining frequent sensitive patterns.

The proposed algorithm has other promising features which are listed below:

- no hiding failure
- very low misses cost
- no false drops
- minimum sanitization rate
- very low proportion of dissimilarity between the source and the sanitized dataset
- linear increase/decrease in execution time
- no elimination of transactions
- single scan of source dataset
- use of simpler data structures

This approach is efficient and scalable in the sense that the execution time is predictable and acceptable in large datasets. Desanitization or reconstruction of the original dataset is not at all possible in this approach, as the alterations to the original dataset are not saved anywhere and no encryption technique is used. This approach does not require any overhead to the mining process after sanitizing a dataset. No adversary or counterpart can mine the hidden patterns even with an arbitrarily low support threshold values.

# References

[1]  Clifton.C & Marks.C, "Security and Privacy Implications of Data Mining", in *Workshop on Data Mining and Knowledge Discovery* -  Montreal, Canada, pp. 15-19.

[2]  Atallah.M, Bertino.E, Elmagarmid.A, Ibrahim.M, and Verykios.V, "Disclosure Limitation of Sensitive Rules",  In *Proc. of IEEE Knowledge and Data Engineering Workshop*, pages 45–52, Chicago, Illinois, November 1999.

[3]  Verykios.V, Gkoulalas-Divanis.A, "A survey of association rule hiding methods for privacy," in *Privacy-Preserving Data Mining: Models and Algorithms*, C. Aggarwal and P. Yu, Eds. New York: Springer, 2008, pp. 267-289.

[4]  Gkoulalas-Divanis.A, Verykios.V, "An integer programming approach for frequent itemset hiding," *Proc. 15th ACM international conference on Information and knowledge management*, 2006, pp. 748-757.

[5]  Gkoulalas-Divanis.A, Verykios.V, "Exact knowledge hiding through database extension," *IEEE Transactions on Knowledge and Data Engineering*, vol.(21, 2008), pp. 699-713.

[6]  Mannila.H, Toivonen.H, "Levelwise search and borders of theories in knowledge discovery," *Data Mining and Knowledge Discovery*, vol.(1), 1997, pp. 241-258.

[7]  Sun.X, Yu.P, "A border-based approach for hiding sensitive frequent itemsets," *Proc. 5th IEEE International Conference on Data Mining*, 2005, pp. 426-433.

[8]  Sun.X, Yu.P, "Hiding sensitive frequent itemsets by a border-based approach," *Journal of Computing Science and Engineering*, vol.(1, 2007), pp. 74-94.

[9]  Mousakides.G, Verykios.V, "A max min approach for hiding frequent itemsets,"  *Data and Knowledge Engineering*, Vol(65, 2008), pp. 75-89.

[10] Oliveira.S, Zaiane.O, "Privacy preserving frequent itemset mining," *Proc2002 IEEE International Conference on Privacy, Security and Data Mining*, 2002, pp. 43-54.

[11] Saygin.Y, Verykios.V, Clifton.C, "Using unknowns to prevent discovery of association rules," *ACM SIGMOD Records,* vol.( 30, 2001), pp. 45-54.

[12] Agrawal R, Srikant R. "Fast algorithms for mining association rules", *Proceedings of 20$^{th}$ international conference on very large data bases*, Santiago, Chile,1994. p.487-99.

[13] Han, J., Pei, J., Yin, Y.: "Mining frequent patterns without candidate generation," *Proc. 2000 ACM SIGMOD international conference on Management of data, 2000*, pp. 1-12.

[14] Pavon.J, Viana.S, Gomez.S, "Matrix Apriori: speeding up the search for frequent patterns," *Proc. 24th IASTED International Conference on Databases and Applications, 2006*, pp. 75-82.

[15] Yildz.B, and Ergenc.B, "Hiding Sensitive Predictive Frequent Itemsets", *Proceedings of the International MultiConference of Engineers and Computer Scientists* 2011, Vol-I.

[16] Cynthia Selvi.P, Mohamed Shanavas.A.R, "An Effective Heuristic Approach for Hiding Sensitive Patterns in Databases", *IOSR-Journal on Computer Engineering,* Volume:Issue(5:1), Sep-Oct, 2012, PP 06-11, DOI. 10.9790/0661-0510611.

[17] Cynthia Selvi.P, Mohamed Shanavas.A.R, "An Improved Item-based Maxcover Algorithm to Protect Sensitive Patterns in Large Databases", IOSR-Journal on Computer Engineering, Volume 14, Issue 4 (Sep-Oct, 2013), PP 01-05, DOI. 10.9790/0661-1440105.

[18] Cynthia Selvi.P, Mohamed Shanavas.A.R, "Introducing Parallelism in Privacy Preserving Data Mining Algorithms"- International Journal of Computational Engineering Research – Vol.4, Issue 2 – Feb 2014, PP. 38 - 41.

[19] The Dataset used in this work for experimental analysis was generated using the generator from IBM Almaden Quest research group and is publicly available from *http://fimi.ua.ac.be/data/*

[20] Ellis Horowitz, Sartaj Sahni, Sanguthevar Rajasekaran, *"Fundamentals of Computer Algorithms"*, Galgotia Pub. Pvt. Ltd, Delhi, 1999.

**Authors' Biography :**

1. **Correspondent Author and Research Scholar : Mrs. P. Cynthia Selvi**

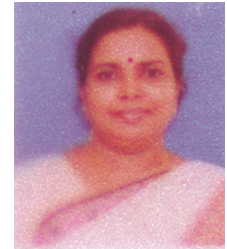

Mrs.P.Cynthia Selvi, qualified with M.Sc., M.Phil has joined Government Collegiate Education Service in Tamilnadu State, India in the year 1991. Now she is working as the Head & Associate Professor in Dept. of Computer Science at Kunthavai Nachiyar Govt. Arts College for Women(Autonomous), Thanjavur-613007, Tamilnadu, India. She has presented 4 papers in International/National Conference/Seminar and has published 5 articles in International Journals. Her areas of interest are Data Mining, Soft Computing Techniques and Embedded Systems. Presently she is a Research scholar in Computer Science in Bharathidasan University, Trichirappalli, Tamilnadu, India and pursuing Ph.D., under the guidance of Research Advisor, Dr. A. R. Mohamed Shanavas.

2. **Co-Author and Research Advisor : Dr. A.R. Mohamed Shanavas**

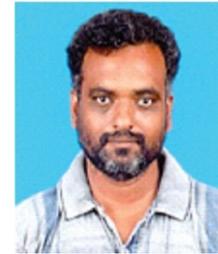

Dr. A. R. Mohamed Shanavas, qualified with M.Sc., PGDCA., M.Ed., M.Phil, Ph.D., Degrees is presently working as Associate Professor in Department of Computer Science, Jamal Mohammed College (Autonomous) affiliated to Bharathidasan University, Trichirapalli-620020, Tamilnadu State, India. Having 24 years of experience in this faculty, he is specialized in the areas of Computer Graphics, Image Processing, Data Mining and Soft Computing Techniques. He has acted as a resource person and chairperson for paper presentation session in International/National conferences. He has Presented papers in various International/National/ State Level Conference/Seminar/Syposium/ Workshop since 2006 and has published 11 articles in International and National journals. 35 candidates were completed the M.Phil candidature under his guidance; 5 and 8 candidates are pursuing M.Phil and Ph.D., respectively. He has Organised a National Conference on "Advanced Concepts in Computer Science" in September 2010. He has visited the Kingdom of Saudi Arabia during May 2013.